\documentclass[conference]{IEEEtran}
\IEEEoverridecommandlockouts
\usepackage{cite}
\usepackage{amsmath,amssymb,amsfonts}
\usepackage{algorithmic}
\usepackage{graphicx}
\usepackage{textcomp}
\usepackage{xcolor}
\usepackage{todonotes} 
\usepackage{url}
\usepackage{amsthm}
\usepackage{multirow}

\usepackage{tablefootnote}
\usepackage{booktabs}
\usepackage{hyperref}
\usepackage{pifont}
\usepackage{balance}
\usepackage{setspace}

\begin{document}
\setstretch{0.95}

\title{A Functional Software Reference Architecture for LLM-Integrated Systems

\thanks{This research work has been funded by the Swedish Knowledge Foundation through the MoDEV project (20200234) , by Vinnova through the iSecure project(202301899), and by the KDT Joint Undertaking through the  MATISSE project (101140216).}}

\author{\IEEEauthorblockN{
Alessio Bucaioni\IEEEauthorrefmark{1},
Martin Weyssow\IEEEauthorrefmark{2}
Junda He\IEEEauthorrefmark{2},
Yunbo Lyu\IEEEauthorrefmark{2},
David Lo\IEEEauthorrefmark{2}
}

\IEEEauthorblockA{\IEEEauthorrefmark{1}
M\"{a}lardalen University (Sweden), \textit{alessio.bucaioni@mdu.se}}
\IEEEauthorblockA{\IEEEauthorrefmark{2}
Singapore Management University (Singapore), \textit{\{mweyssow,jundahe,yunbolyu,davidlo\}@smu.edu.sg}}
\vspace{-1.cm}
}

\maketitle



\begin{abstract}

The integration of large language models into software systems is transforming capabilities such as natural language understanding, decision-making, and autonomous task execution. However, the absence of a commonly accepted software reference architecture hinders systematic reasoning about their design and quality attributes. This gap makes it challenging to address critical concerns like privacy, security, modularity, and interoperability, which are increasingly important as these systems grow in complexity and societal impact. In this paper, we describe our \textit{emerging} results for a preliminary functional reference architecture as a conceptual framework to address these challenges and guide the design, evaluation, and evolution of large language model-integrated systems. We identify key architectural concerns for these systems, informed by current research and practice. We then evaluate how the architecture addresses these concerns and validate its applicability using three open-source large language model-integrated systems in computer vision, text processing, and coding
\end{abstract}


\vspace{-0.2cm}
\section{Introduction}
\vspace{-0.1cm}

The integration of large language models (LLMs) into software systems is transforming the software landscape, enabling capabilities like natural language understanding and autonomous task execution~\cite{achiam2023gpt}. LLMs are driving innovation across domains such as customer service, healthcare, and education~\cite{bommasani2021opportunities, klarna_ai_assistant}. However, serving pre-trained or fine-tuned models to end-users introduces complexities, including modularity and reuse~\cite{xiao2024configurable}, quality attribute trade-offs~\cite{shashidhar2023democratizing}, and challenges like bias, fairness, regulatory compliance~\cite{eu_ai_act}, and ethical use.
Software architecture plays a key role in shaping system quality~\cite{bass2012software}. Reference architectures (RAs) provide standardized guidance for design, reuse, and evolution~\cite{garces2021three, ferko2023standardisation}. Well-known RAs, such as AUTOSAR and the NIST Big Data Framework, have reduced costs, improved interoperability, and promoted best practices~\cite{garces2021three}. More recently, the digital twin domain has highlighted their value for emerging fields~\cite{garces2021three}.
Despite the rapid adoption of LLM-integrated systems designed to serve inference capabilities to end-users, no documented RAs exist to address their design and quality attributes. This gap leaves stakeholders without systematic guidance to tackle critical concerns, hindering design reuse and scalability. As these systems grow in complexity and societal impact, the need for a guiding framework becomes increasingly urgent.

In this paper, we present our \textit{emerging} results for a preliminary functional reference architecture to address the challenges of LLM-integrated systems. We identify and analyze key architectural concerns for systems serving pre-trained or fine-tuned models for inference tasks, then propose a preliminary functional software RA~\cite{garces2021three, behere2016functional}. Our work is grounded in a review of software architecture literature for LLMs and our experience in LLM research and applications~\cite{yang2024robustness, nazir2024architecting}. Finally, we evaluate the proposed architecture against the identified concerns and validate its applicability through three open-source LLM-integrated systems in computer vision, text processing, and coding.

The remainder of this paper is organized as follows.
Section~\ref{sec:research_methodology} presents the architectural concerns for LLM-integrated systems.
Section~\ref{sec:ra} details the proposed reference architecture.
Section~\ref{sec:val} validates the architecture against the identified concerns and analyses three open-source LLM-integrated systems.
Finally, Section~\ref{sec:discussion} concludes with a discussion and outlines directions for future research.

\vspace{-0.1cm}
\section{Architectural concerns}\label{sec:research_methodology}\vspace{-0.1cm}

This section presents the architectural concerns for LLM-integrated systems, identified through a targeted literature review and enriched by established architectural principles and extensive experience in LLM research and applications~\cite{yang2024robustness, nazir2024architecting}.

We searched IEEE Xplore, ACM Digital Library, and Scopus~\cite{kitchenham2013systematic} using a concise query string applied to titles and abstracts:
\textit{“software architecture” AND (“LLM*” OR "large language model*”)}.
The search yielded 38 peer-reviewed publications, refined to 26 after removing duplicates and irrelevant entries. Applying inclusion criteria focused on peer-reviewed studies addressing LLM software architectures~\cite{ali2014evaluating}, the set was narrowed to two studies. To mitigate biases~\cite{Greenhalgh:2005}, backward and forward snowballing~\cite{snowball} identified one additional study, resulting in three primary studies~\cite{paper1, paper2, paper3} \footnote{The interested reader can refer to the automatic search and selection process detailed in~\cite{slr}}. These studies were analysed and further enriched with insights from architectural principles and practical experience, leading to the concerns summarized in Table~\ref{tab:concerns}. It is important to note that we do not claim these to be the only or exhaustive architectural concerns for such systems.

\begin{table}[]
\scriptsize{
\begin{center}{
\caption{Architectural concerns}
\vspace{-0.1cm}
\begin{tabular}{| p{1.5cm} | p{6.5cm}|}
   \hline
   \textbf{Concern} & \textbf{Description}\\
	\hline\hline 
LLM integration & Incorporating LLMs into software systems involves exposing their capabilities as reusable services. \\ \hline 
Data handling & LLMs require efficient data pipelines and handling of structured and unstructured data. \\ \hline 
User interaction & Interfaces should provide seamless access while abstracting complexity. \\ \hline
Performance &  Minimizing latency during inference while maintaining accuracy. \\ \hline
Scalability & Handle increasing loads (e.g., concurrent queries, larger datasets). \\ \hline
Security &  Protecting against threats like unauthorized access, malicious prompts, and data breaches. \\ \hline
Privacy and compliance & Adherence to data privacy regulations (e.g., GDPR, EU AI Act). \\ \hline
Modularity & Breaking down LLMs functionalities into reusable components. \\ \hline 
Customizability & Customizing LLMs for domain-specific tasks through prompt engineering or retrieval-based methods. \\ \hline
Interoperability &  Ensuring seamless integration of LLMs with existing tools, databases, and APIs. \\ \hline
Fairness & Addressing biases to ensure equitable and ethical usage.\\ \hline
   \end{tabular}}
   \label{tab:concerns}
\end{center}
\vspace{-0.7cm}}
\end{table}

\textit{LLM integration} requires exposing model capabilities as reusable services to avoid monolithic embedding, enabling modularity, scalability, and dynamic updates. For instance, OpenAI’s GPT services leverage RESTful APIs for modular access.
\textit{Data handling} and \textit{performance} are critical for managing structured and unstructured data efficiently while minimizing latency and maintaining reliability. Optimized architectures using caching, parallelization, and hardware acceleration, as demonstrated by HuggingFace’s Transformers library~\cite{wolf2020transformers}, ensure low-latency inference and efficient resource utilization.
\textit{Scalability} addresses the challenges of large model sizes and high concurrent query volumes. Solutions like OpenAI’s GPT-4~\cite{achiam2023gpt} APIs use distributed clusters and horizontal scaling to manage millions of daily requests, ensuring robust performance under load.
\textit{User interaction} focuses on seamless, intuitive interfaces that abstract LLM complexity while providing mechanisms for feedback and control over probabilistic outputs. Microsoft’s GitHub Copilot exemplifies this through its integration with IDEs, enhancing usability and developer productivity.
\textit{Security} and \textit{privacy} are essential to safeguard against threats such as unauthorized access, malicious prompts, and data breaches. Measures like input validation, encryption, and secure deployment pipelines are critical, with OWASP highlighting risks like prompt injection. Federated learning, as used by Google on Android devices, ensures compliance with privacy regulations like GDPR by minimizing data exposure.
\textit{Modularity} simplifies updates and enhances scalability by breaking LLM functionalities into reusable components. AWS Sagemaker exemplifies this with modular deployments of preprocessing and inference components, enabling flexible system maintenance and scaling.
\textit{Customizability} supports domain-specific adaptations without disrupting core architecture through methods like prompt engineering and fine-tuning. OpenAI’s GPT fine-tuning API allows tailoring models for use cases such as customer support and legal text generation.
\textit{Interoperability} ensures seamless integration with existing tools, databases, and APIs, avoiding adoption barriers. Salesforce’s Einstein GPT demonstrates this by using adapters to interact smoothly with CRMs and workflows.
\textit{Fairness} focuses on mitigating biases to ensure equitable and ethical usage. Google’s AI ethics team addresses this through bias detection and mitigation strategies during model training, promoting fairness and inclusivity.

\vspace{-0.1cm}
\section{Proposed software reference architecture} \label{sec:ra}
\vspace{-0.1cm}
In this section, we present the preliminary functional software reference architecture (RA) for LLM-integrated systems. Following Garces et al.\cite{garces2021three}, preliminary indicates that the RA synthesizes insights from existing systems while proposing generalized principles for future design. Based on Behere and Törngren\cite{behere2016functional}, functional emphasizes a black-box view of the system, detailing entities and interactions. This aligns with the functional view in ISO/IEC/IEEE 42010:2022, which standardizes the architectural description of software-intensive systems~\cite{ISO_IEC_20233_2019}.
\vspace{-0.2cm}
\begin{figure}[h!]
\centering
    \includegraphics[width=\linewidth]{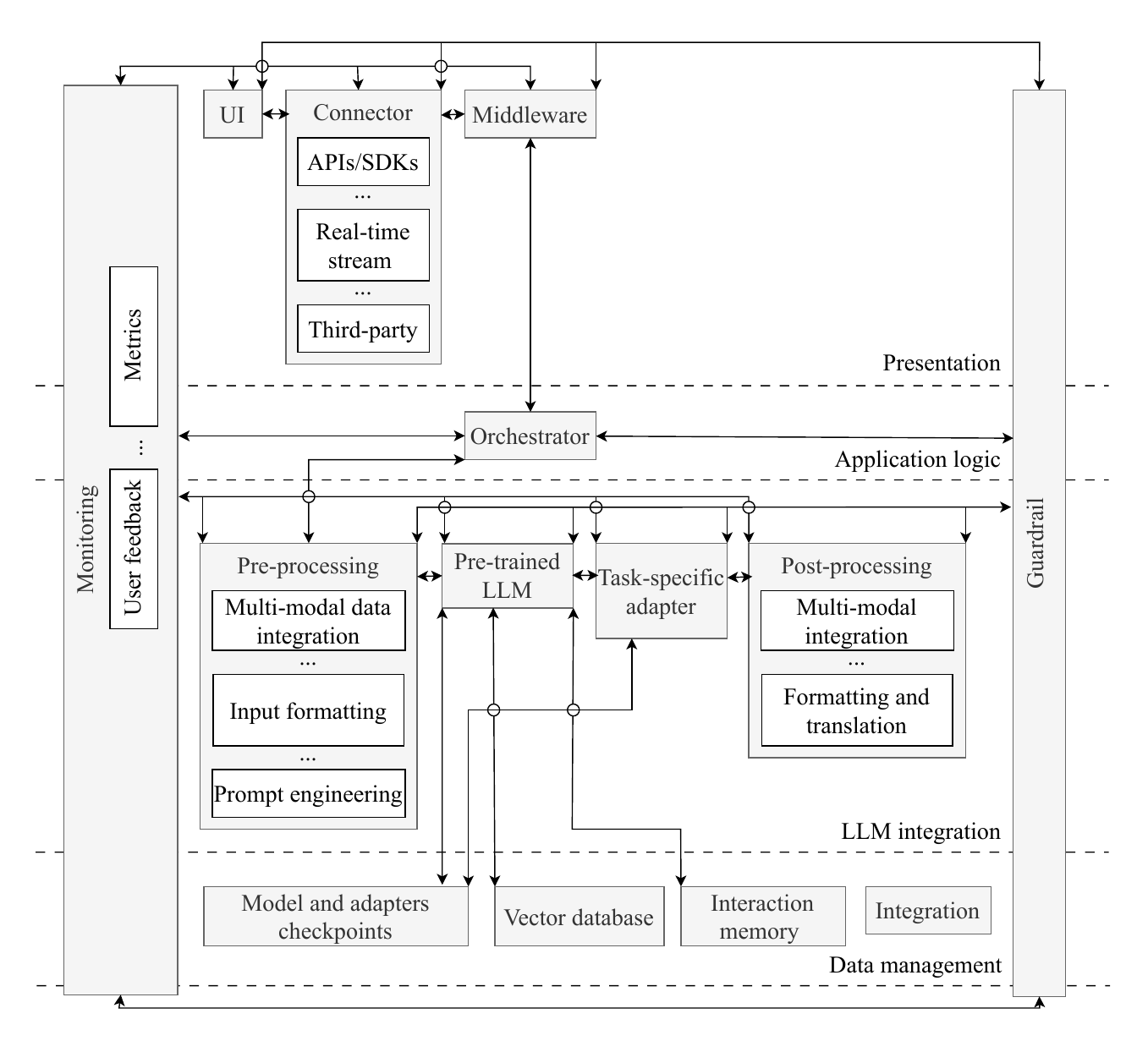}
    \caption{Preliminary functional software RA for LLMs-integrated Systems}
    \label{fig:ra}
       \vspace{-.1cm}
\end{figure}

Figure~\ref{fig:ra} shows the RA, organized into four layers: \textit{Presentation}, \textit{Application logic}, \textit{LLM integration}, and \textit{Data management}, represented by dotted black lines. Each layer contains functional components (grey boxes) and optional sub-components (white boxes), representing example functionalities with ellipses indicating possible extensions. Two sidecars—\textit{Monitoring} and \textit{Guardrail}—span multiple layers. Connections are shown with directional arrows, and overlapping connections are marked by black-bordered white circles for clarity. Each component operates as an independent service to enable modular updates and scaling of LLM functionalities.

The Presentation layer facilitates user interaction and visualization, acting as the entry point for external systems and users. It includes a \textit{UI} component for interfaces like web apps, mobile applications, and chatbots, such as ChatGPT’s web interface. The \textit{Connector} bridges LLMs with external services, supporting real-time streams (e.g., WebSocket) and third-party integrations, such as Slack’s integration with ChatGPT for task management. The \textit{Middleware} handles request validation, transformation, and logging, maintaining conversational context similar to OpenAI’s Session Management Layer. Communication within this layer typically uses secured HTTP/HTTPS or GraphQL, while event-driven middleware ensures responsive interactions.
The Application logic layer manages task and data flow within the system. Its core component, the \textit{Orchestrator}, dynamically determines workflows based on user inputs, such as summarization or translation. For example, Jasper AI’s Content Workflow demonstrates this functionality. The Orchestrator supports asynchronous, event-driven workflows to ensure efficient execution and scalability. 
The LLM integration layer is the system's core, handling input and output processing. The \textit{Pre-processing} component formats and enriches inputs with tasks like tokenization, prompt engineering, and multi-modal integration, as seen in Google’s Vertex AI Pipelines. The central inference relies on the \textit{Pre-trained LLM} and \textit{Task-specific adapters} for fine-tuned, domain-specific responses, such as ChatGPT’s adapters for FAQs or personalized support. The \textit{Post-processing} component refines outputs (e.g., formatting, translation) for user-friendly delivery, exemplified by DALL-E’s Caption Formatter, which converts image outputs into natural language descriptions. Components communicate via RESTful APIs, gRPC for low latency, or shared caching mechanisms.
The Data management layer ensures efficient data handling. \textit{Model and adapter checkpoints} store pre-trained models and fine-tuned adapters for seamless updates. A \textit{vector database} enables retrieval-augmented generation (RAG) for knowledge-grounded outputs, improving accuracy, as seen in Pinecone’s integration with Notion AI. The \textit{Interaction memory} component maintains context across sessions. Long-term memory stores persistent user data (e.g., profiles, preferences, and error history), while short-term memory retains session-specific chat context, as exemplified by ChatGPT’s Memory Feature. \textit{Integration} components connect with external APIs, knowledge bases, and tools, ensuring system interoperability.
The \textit{Monitoring} sidecar spans all layers, collecting metrics like latency, throughput, and system health, along with user feedback. For example, Azure OpenAI’s Metrics Dashboard monitors inference latency, token usage, and error rates to identify bottlenecks and improve performance.
The \textit{Guardrail} sidecar enforces security and compliance across layers, validating user inputs and ensuring adherence to regulations like GDPR and the EU AI Act. For instance, Microsoft Copilot’s Compliance Checks prevent outputs containing confidential or inappropriate content.

Intra-component communication ensures seamless interaction across all layers. The Presentation layer communicates with the Application logic layer via secured HTTP/REST or GraphQL calls, returning responses in JSON. The Application logic layer connects to the LLM integration layer through synchronous (REST/GraphQL) and asynchronous (e.g., RabbitMQ, Kafka) methods for real-time tasks and batch processing. The LLM integration layer exposes APIs through REST or gRPC, with outputs serialized as JSON or Protocol Buffers (Protobuf) for high performance. The Data management layer provides data to the LLM integration layer via REST APIs or database queries, storing embeddings, interaction histories, and structured data for context-aware operations.
The Monitoring sidecar collects metrics asynchronously using message brokers like Kafka and aggregates them into a central system. Alerts and anomalies are communicated back to the Application logic layer via REST APIs or notification services. The Guardrail sidecar validates inputs inline at the Presentation layer and enforces security through synchronous REST APIs or policy hooks.

\vspace{-0.2cm}
\section{Validating the proposed reference architecture} \label{sec:val}
\vspace{-0.1cm}
This section evaluates how the proposed RA addresses the identified concerns and validates its applicability through three open-source LLM-integrated systems in computer vision, text processing, and coding.

\subsection{Proposed reference architecture Vs. architectural concerns}
Table~\ref{tab:val1} demonstrates how the proposed reference architecture (RA) addresses key architectural concerns through a modular, layered design with cross-layer components.
\vspace{-0.2cm}
\begin{table}[h!]
\scriptsize{
\begin{center}{
\caption{Proposed reference architecture Vs. architectural concerns}
\vspace{-0.1cm}
\begin{tabular}{| p{1.7cm} | p{6.2cm}|}
   \hline
   \textbf{Concern} & \textbf{How the concern is satisfied}\\
	\hline\hline 
LLM integration & The LLM integration layer encapsulates LLM functionalities as modular services, accessible via APIs for easy integration and independent updates. \\ \hline 
Data handling & Data pre-processors ensure efficient formatting and batching; adapters integrate external data sources; vector databases support retrieval-augmented generation. \\ \hline 
User interaction & The Presentation layer provides user-facing interfaces and unified access to underlying services through API gateways, ensuring smooth interactions and validation. \\ \hline
Performance and scalability &  Distributed processing and asynchronous workflows optimize performance under high loads; auto-scaling and caching ensure efficient handling of concurrent requests. \\ \hline
Security, privacy, compliance and fairness & Secure communication with TLS; authentication managed at Integrator; Guardrail enforces compliance policies, logs, and mitigates risks like malicious prompts and breaches. \\ \hline
Modularity & Layers separate concerns; microservices allow independent updates and scaling of functionalities. \\ \hline 
Customizability & Fine-tuning and inference pipelines are modular, enabling domain-specific adaptations and reusable components without disrupting the core architecture.\\ \hline
Interoperability &  Standardized APIs and integration adapters ensure seamless interaction with external systems, databases, and workflows. \\ \hline
   \end{tabular}}
   \label{tab:val1}
\end{center}
\vspace{-0.3cm}
}
\end{table}

\begin{table*}[htbp!]
\scriptsize{
\begin{center}{
\caption{Proposed reference architecture Vs. existing LLM-integrated systems}
\vspace{-0.1cm}
\small
\begin{tabular}{| p{2.6cm} | p{5cm}| p{4.5cm}| p{4.3cm}|}
   \hline
   \textbf{Component} & \textbf{MaxKB} & \textbf{Continue} & \textbf{InternVL}\\
	\hline\hline 
    UI & UI in Vue.js & Continue Sidebar, Chat, In-editor UI & UI in Streamlit Python	\\\hline
    Connector & RESTful APIs, LangChain Connectors & VSCode API, Web View, LLM Providers APIs & RESTful APIs \\\hline
    Middleware & Request Validation, Input Transformation in Django & -- & Request Validation, Input Transformation\\\hline
    Orchestrator & Workflow Engine & VSCode Extension & Workflow Engine \\\hline
    Pre-processing & Context Retrieval, Prompt Reformulation, Tokenization, Image Transformation in Langchain & Context Retrieval and Prompt Reformulation & Caption Generation, Image Transformation, Tokenization \\\hline
    Pre-trained LLM & Multiple LLMs & Multiple LLMs &  Multiple LLMs \\\hline
    Task-specific adapter & -- & -- & LoRA \\\hline
    Post-processing & Output Parsing, Filtering, and  Formatting in LangChain & Output Filtering, Code Change Integration &  Output parsing (Multi-modality), Output Formatting \\\hline
    Model and adapters checkpoints & Load LLMs via LangChain or Locally via Ollama & Load LLMs from providers'APIs or Locally via Ollama & Load LLMs from providers' APIs \\\hline
    Vector Database & Knowledge Vectors with pgvector & Codebase Vectors & --\\\hline
    Interaction Memory & Session Storage, Conversational Memory (PostgreSQL)
 & Cache Completion & Session Storage, Conversational Memory  \\\hline
    Integration & RESTful APIs, Customized Functions \& tools & RESTful API (e.g., OpenAI, Anthropic, Ollama) & RESTful API (e.g., OpenAI-compatible) \\\hline
    Monitoring & System Usage, User Feedback & Telemetry (PostHog), User Feedback & User Feedback \\\hline
    Guardrail & -- & -- & Safeguard for content moderation \\\hline
   \end{tabular}}
   \label{tab:val2}
\end{center}
\vspace{-0.6cm}
}
\end{table*}

LLM integration is handled by the LLM integration layer, which modularizes pre-processing, inference, and post-processing using microservices and proxy patterns. These components operate independently, accessible via standardized APIs, ensuring seamless integration, scalability, and maintainability.
Data handling is supported by the data management layer, which uses vector databases for retrieval-augmented generation (RAG) and embedding-based operations, enabling low-latency access to external knowledge. The interaction memory component maintains session context, enhancing personalization and enabling context-aware user interactions.
User interaction is addressed in the presentation layer, which provides adaptable interfaces, such as web apps and chatbots, to abstract LLM complexity. Integrators and middleware ensure smooth, real-time communication with external systems by managing request validation, transformation, and session handling.
Performance and scalability are achieved through event-driven communication, distributed processing, and horizontal scaling. Asynchronous workflows optimize resource utilization, while caching mechanisms and task-specific optimizations improve response times for high-throughput applications.
The Monitoring and Guardrail sidecars address security, compliance, and system optimization. Monitoring collects runtime metrics like latency and error rates, enabling real-time performance tracking and proactive improvements. Guardrails enforce compliance with privacy regulations (e.g., GDPR) and mitigate risks such as malicious prompts, ensuring secure and ethical operations.
Fairness and privacy are reinforced through bias detection and regulatory compliance enforced by the Guardrail sidecar. Monitoring further supports fairness by detecting anomalies and enabling continuous improvements.
Modularity and interoperability are strengths of the LLM integration layer, where independent pre-processing, inference, and post-processing components allow seamless updates and domain-specific adaptations, such as fine-tuning or prompt engineering, without affecting the core system.
Customizability is achieved via modular task-specific adapters and fine-tuned models, enabling tailored outputs for diverse domains while preserving system stability, efficiency, and reliability.

\subsection{Proposed reference architecture Vs. existing LLM-integrated systems}

Table~\ref{tab:val2} maps three open-source LLM-integrated systems—spanning text processing, coding, and computer vision—to the proposed preliminary functional software RA.

Max Knowledge Base (MaxKB) combines LLMs with external knowledge retrieval to deliver accurate, domain-specific responses~\cite{MaxKB}. It provides a Vue.js-based user interface for interaction and supports workflow orchestration through drag-and-drop components. MaxKB integrates with third-party tools for dynamic task execution and uses PostgreSQL with pgvector to store document embeddings for RAG. Documents are segmented and stored as embeddings to enable efficient retrieval and Dialogue logs maintain session histories and user feedback, enabling query analysis over time ranges (e.g., past 7, 30, or 180 days) for refining AI responses. Built on LangChain~\cite{Chase_LangChain_2022}, MaxKB integrates multiple LLM providers, such as Llama 3, Qwen 2, OpenAI, Claude, and dynamically combines top-k retrieved segments into contextually relevant prompts, aligning with the LLM Integration Layer for pre-processing and inference.

Continue, a coding assistant integrated with VSCode and boasting nearly 20,000 GitHub stars~\cite{continue-plugin}, exemplifies modular LLM integration. It offers three interfaces: a sidebar for configuration, a chat for LLM interactions, and in-editor tools for tasks like code autocompletion and bug fixing. Continue connects to LLM providers (e.g., OpenAI, Anthropic, Google Gemini) and supports local LLMs via Ollama. The Orchestrator in the Application Logic Layer activates workflows like docstring generation or code autocompletion. 
The plugin indexes the project codebase into a vector database that is leveraged by a context retrieval module to augment user prompts with contextually relevant code.
It supports multiple LLMs for code autocompletion, chat, code editing, and code retrieval.
LLMs outputs are filtered to remove redundancy before merging accepted changes into the codebase, aligning with the RA’s Post-processing. Telemetry powered by PostHog tracks user interactions, while feedback mechanisms allow issue reporting, aligning with the Monitoring Sidecar.

InternVL supports multi-modal tasks such as text generation, reasoning, and dialogue, processing images, videos, and text~\cite{internvl2024}. Its Streamlit-based user interface allows local and web-based interaction, while RESTful APIs support external integrations, mapping to the Presentation Layer. The middleware handles input validation, transformation (e.g., tokenization, image resizing), and logging, ensuring session context for multi-turn conversations. The Orchestrator dynamically selects workflows for text, images, or videos, enabling tasks like captioning, multi-image reasoning, and video processing. The Pre-processing component handles tokenization and image transformations, while Post-processing refines outputs for integration.
InternVL uses pre-trained multimodal LLMs (e.g., InternVL2-76B) with support for fine-tuning, including parameter-efficient methods like LoRA, aligning with the LLM Integration Layer. Interaction memory preserves session context for multi-turn conversations, enhancing personalization. LMDeploy enables OpenAI-compatible API deployment, supporting interoperability. Telemetry tracks user interactions (e.g., upvotes, flags) for performance improvement, while Guardrails enforce content moderation, ensuring outputs adhere to ethical and safety guidelines.

\section{Final Discussion and Future Steps}\label{sec:discussion}
\vspace{-0.05cm}

Compared to existing work, our study provides a novel contribution by proposing a preliminary functional software RA for LLM-integrated systems. Unlike related work, such as foundation model-based systems~\cite{paper2}, our RA explicitly addresses unique architectural concerns of LLM-driven systems, including modular pre-processing, dynamic task orchestration, and cross-layer monitoring.
While our work is grounded in software architecture literature, established principles, and extensive experience in LLM research, we acknowledge limitations in the scope and systematicity of our review. However, given the limited availability of peer-reviewed studies on LLM-integrated systems, this is unlikely to introduce significant bias, as additional findings would more likely complement rather than challenge our proposed RA.
The proposed RA synthesizes insights from existing systems while also proposing generalized principles for future LLM-integrated system designs. It is important to note that current systems may not necessarily include all the identified RA components (e.g., Task-specific adapters and Guardrail components), which is typical for RAs, as demonstrated in previous research on domains like DTs~\cite{ferko2023standardisation}. Additionally, there may be architectural concerns not yet considered, which is a natural aspect of the early-stage development of such frameworks. We anticipate that further research and practical applications will help identify and address these gaps.
To address the above limitations, we are conducting a systematic literature review, including grey literature, to capture broader perspectives and emerging insights into this rapidly evolving field.
The proposed RA was validated against three open-source LLM-driven systems—MaxKB, Continue, and InternVL. However, further evaluation across real-world systems spanning diverse domains is essential to ensure broader coverage and uncover any overlooked challenges. As part of future work, we also plan to conduct focus group validations and expert surveys to gather qualitative feedback on the RA’s design and its practical applicability in integrated system architectures.
We also aim to extend this RA by incorporating dedicated architectural views, as suggested by the ISO/IEC/IEEE 42010 standard, such as the technical view, and views addressing unique concerns of LLM-integrated systems, including compliance and fairness. This extension may ultimately form a comprehensive architectural framework, similar to those introduced for the automotive domain~\cite{pelliccione2017automotive}.
\vspace{-0.2cm}

\balance
\bibliographystyle{IEEEtranS}

\bibliography{main}

\begin{thebibliography}{10}
\providecommand{\url}[1]{#1}
\csname url@samestyle\endcsname
\providecommand{\newblock}{\relax}
\providecommand{\bibinfo}[2]{#2}
\providecommand{\BIBentrySTDinterwordspacing}{\spaceskip=0pt\relax}
\providecommand{\BIBentryALTinterwordstretchfactor}{4}
\providecommand{\BIBentryALTinterwordspacing}{\spaceskip=\fontdimen2\font plus
\BIBentryALTinterwordstretchfactor\fontdimen3\font minus
  \fontdimen4\font\relax}
\providecommand{\BIBforeignlanguage}[2]{{%
\expandafter\ifx\csname l@#1\endcsname\relax
\typeout{** WARNING: IEEEtranS.bst: No hyphenation pattern has been}%
\typeout{** loaded for the language `#1'. Using the pattern for}%
\typeout{** the default language instead.}%
\else
\language=\csname l@#1\endcsname
\fi
#2}}
\providecommand{\BIBdecl}{\relax}
\BIBdecl

\bibitem{continue-plugin}
``{Continue}, open-source ai code assistant.''
  \url{https://github.com/continuedev/continue}, accessed: 2024-12-09.

\bibitem{MaxKB}
1Panel-dev, ``Maxkb: An open-source knowledge base q\&a system,''
  \url{https://github.com/1Panel-dev/MaxKB}, 2024, accessed: 2024-12-12.

\bibitem{achiam2023gpt}
J.~Achiam, S.~Adler, S.~Agarwal, L.~Ahmad, I.~Akkaya, F.~L. Aleman, D.~Almeida,
  J.~Altenschmidt, S.~Altman, S.~Anadkat \emph{et~al.}, ``Gpt-4 technical
  report,'' \emph{arXiv preprint arXiv:2303.08774}, 2023.

\bibitem{ali2014evaluating}
N.~B. Ali and K.~Petersen, ``Evaluating strategies for study selection in
  systematic literature studies,'' in \emph{Procs of ESEM}, 2014.

\bibitem{bass2012software}
L.~Bass, \emph{Software architecture in practice}.\hskip 1em plus 0.5em minus
  0.4em\relax Pearson Education India, 2012.

\bibitem{behere2016functional}
S.~Behere and M.~T{\"o}rngren, ``A functional reference architecture for
  autonomous driving,'' \emph{Information and Software Technology}, vol.~73,
  pp. 136--150, 2016.

\bibitem{bommasani2021opportunities}
R.~Bommasani, D.~A. Hudson, E.~Adeli, R.~Altman, S.~Arora, S.~von Arx, M.~S.
  Bernstein, J.~Bohg, A.~Bosselut, E.~Brunskill \emph{et~al.}, ``On the
  opportunities and risks of foundation models,'' \emph{arXiv preprint
  arXiv:2108.07258}, 2021.

\bibitem{slr}
{Bucaioni et al.}, ``Automatic search, and selection,''
  \url{https://docs.google.com/spreadsheets/d/1Tm3m4iMFQsZMHGfhCH6NjhbIq_UKyk8T/edit?usp=share_link},
  accessed: 2024-12-02.

\bibitem{Chase_LangChain_2022}
\BIBentryALTinterwordspacing
H.~Chase, ``{LangChain},'' Oct. 2022. [Online]. Available:
  \url{https://github.com/langchain-ai/langchain}
\BIBentrySTDinterwordspacing

\bibitem{eu_ai_act}
{European Union}, ``The artificial intelligence act,''
  \url{https://artificialintelligenceact.eu}, accessed: 2024-12-02.

\bibitem{ferko2023standardisation}
E.~Ferko, A.~Bucaioni, P.~Pelliccione, and M.~Behnam, ``Standardisation in
  digital twin architectures in manufacturing,'' in \emph{2023 IEEE 20th
  International Conference on Software Architecture (ICSA)}.\hskip 1em plus
  0.5em minus 0.4em\relax IEEE, 2023, pp. 70--81.

\bibitem{garces2021three}
L.~Garc{\'e}s, S.~Mart{\'\i}nez-Fern{\'a}ndez, L.~Oliveira, P.~Valle, C.~Ayala,
  X.~Franch, and E.~Y. Nakagawa, ``Three decades of software reference
  architectures: A systematic mapping study,'' \emph{Journal of Systems and
  Software}, vol. 179, p. 111004, 2021.

\bibitem{Greenhalgh:2005}
T.~Greenhalgh and R.~Peacock, ``Effectiveness and efficiency of search methods
  in systematic reviews of complex evidence: audit of primary sources,''
  \emph{BMJ}, vol. 331, no. 7524, pp. 1064--1065, 2005.

\bibitem{paper2}
T.~H{\"a}ndler, ``A taxonomy for autonomous llm-powered multi-agent
  architectures.'' in \emph{KMIS}, 2023, pp. 85--98.

\bibitem{ISO_IEC_20233_2019}
\BIBentryALTinterwordspacing
\emph{{ISO/IEC 20233:2019 - Information technology -- Cloud computing --
  Interoperability and portability}}, {International Organization for
  Standardization} Std., Oct. 2019, accessed: 2024-12-08. [Online]. Available:
  \url{https://www.iso.org/standard/74393.html}
\BIBentrySTDinterwordspacing

\bibitem{kitchenham2013systematic}
B.~Kitchenham and P.~Brereton, ``A systematic review of systematic review
  process research in software engineering,'' \emph{Information and software
  technology}, 2013.

\bibitem{klarna_ai_assistant}
{Klarna}, ``{Klarna AI Assistant Handles Two-Thirds of Customer Service Chats
  in Its First Month},''
  \url{https://www.klarna.com/international/press/klarna-ai-assistant-handles-two-thirds-of-customer-service-chats-in-its-first-month/},
  accessed: 2024-12-02.

\bibitem{paper1}
Q.~Lu, L.~Zhu, X.~Xu, Y.~Liu, Z.~Xing, and J.~Whittle, ``A taxonomy of
  foundation model based systems through the lens of software architecture,''
  in \emph{Proceedings of the IEEE/ACM 3rd International Conference on AI
  Engineering-Software Engineering for AI}, 2024, pp. 1--6.

\bibitem{paper3}
Q.~Lu, L.~Zhu, X.~Xu, Z.~Xing, S.~Harrer, and J.~Whittle, ``Towards responsible
  generative ai: A reference architecture for designing foundation model based
  agents,'' in \emph{2024 IEEE 21st International Conference on Software
  Architecture Companion (ICSA-C)}.\hskip 1em plus 0.5em minus 0.4em\relax
  IEEE, 2024, pp. 119--126.

\bibitem{nazir2024architecting}
R.~Nazir, A.~Bucaioni, and P.~Pelliccione, ``Architecting ml-enabled systems:
  Challenges, best practices, and design decisions,'' \emph{Journal of Systems
  and Software}, vol. 207, p. 111860, 2024.

\bibitem{internvl2024}
\BIBentryALTinterwordspacing
OpenGVLab, ``Internvl,'' 2024, accessed: 2024-12-12. [Online]. Available:
  \url{https://github.com/OpenGVLab/InternVL}
\BIBentrySTDinterwordspacing

\bibitem{pelliccione2017automotive}
P.~Pelliccione, E.~Knauss, R.~Heldal, S.~M. {\AA}gren, P.~Mallozzi,
  A.~Alminger, and D.~Borgentun, ``Automotive architecture framework: The
  experience of volvo cars,'' \emph{Journal of systems architecture}, vol.~77,
  pp. 83--100, 2017.

\bibitem{shashidhar2023democratizing}
S.~Shashidhar, A.~Chinta, V.~Sahai, Z.~Wang, and H.~Ji, ``Democratizing llms:
  An exploration of cost-performance trade-offs in self-refined open-source
  models,'' \emph{arXiv preprint arXiv:2310.07611}, 2023.

\bibitem{snowball}
C.~Wohlin, ``Guidelines for snowballing in systematic literature studies and a
  replication in software engineering,'' in \emph{Procs of EASE}.\hskip 1em
  plus 0.5em minus 0.4em\relax ACM, 2014, pp. 38:1--38:10.

\bibitem{wolf2020transformers}
T.~Wolf, L.~Debut, V.~Sanh, J.~Chaumond, C.~Delangue, A.~Moi, P.~Cistac,
  T.~Rault, R.~Louf, M.~Funtowicz \emph{et~al.}, ``Transformers:
  State-of-the-art natural language processing,'' in \emph{Proceedings of the
  2020 conference on empirical methods in natural language processing: system
  demonstrations}, 2020, pp. 38--45.

\bibitem{xiao2024configurable}
C.~Xiao, Z.~Zhang, C.~Song, D.~Jiang, F.~Yao, X.~Han, X.~Wang, S.~Wang,
  Y.~Huang, G.~Lin \emph{et~al.}, ``Configurable foundation models: Building
  llms from a modular perspective,'' \emph{arXiv preprint arXiv:2409.02877},
  2024.

\bibitem{yang2024robustness}
Z.~Yang, Z.~Sun, T.~Z. Yue, P.~Devanbu, and D.~Lo, ``Robustness, security,
  privacy, explainability, efficiency, and usability of large language models
  for code,'' \emph{arXiv preprint arXiv:2403.07506}, 2024.

\end{thebibliography}

\end{document}